\documentclass[twocolumn,preprintnumbers,superscriptaddress,unsortedaddress]{revtex4-2}
\usepackage{makeidx}
\usepackage{amssymb}
\usepackage{amsmath}
\usepackage{graphicx}
\usepackage{epstopdf}
\usepackage{dcolumn}
\usepackage{bm}
\usepackage{xcolor}
\usepackage{amsfonts}
\usepackage{units}

\graphicspath{{C:/Users/Eyal/MyDocuments/latex/Texfiles/Diamond/FerriMO/POSA/}}

\newcommand{\func}[1]{\operatorname{#1}}

\begin{document}

\title{Polarimeter optical spectrum analyzer}
\author{Eyal Buks}
\affiliation{Andrew and Erna Viterbi Department of Electrical Engineering, Technion,
Haifa 32000 Israel}
\date{\today }

\begin{abstract}
A coherent optical spectrum analyzer is integrated with a
rotating quarter wave plate polarimeter. The combined polarimeter optical
spectrum analyzer (POSA) allows the extraction of the state of polarization
with high spectral resolution. POSA is used in this work to study two
optical systems. The first is an optical modulator based on a
ferrimagnetic sphere resonator. POSA is employed to explore the
underlying magneto-optical mechanism responsible for modulation sideband
asymmetry. The second system under study is a cryogenic fiber
loop laser, which produces an unequally spaced optical comb. Polarization
measurements provide insights on the nonlinear processes responsible for
comb creation. Characterizations extracted from POSA data provide guidelines
for performance optimization of applications based on these systems under
study.
\end{abstract}

\maketitle

\section{Introduction}

Optical state of polarization (SOP) can be measured using a
variety of techniques \cite{goldstein2017polarized}. For some applications,
the dependency of SOP on optical wavelength $\lambda $ has to be determined. This dependency on $\lambda $ can be obtained by combining a wavelength
tunable band-pass optical filter with the polarimeter being used to measure
the SOP. This relatively simple configuration has a spectral resolution that
is limited by the ratio between the filter's free spectral range (FSR) and its finesse. Lowering the spectral resolution below
about $1\func{GHz}$ over a wide tuning range in the optical band is
practically challenging. On the other hand, for some applications, a much
higher spectral resolution is needed.

A coherent optical spectrum analyzer (OSA) \cite{Yao_17854,Kim_351,Kim_317}
is an instrument based on heterodyne detection of the optical signal under
study \cite{Baney_355,Feng_348,Dang_4430}. Commonly, a coherent OSA has
significantly higher spectral resolution compared to grating-based OSA. In
its standard configuration, SOP cannot be extracted from coherent OSA data. %
In this work a fiber-based system is proposed, which integrates a coherent OSA with a polarimeter (PM) that is based on a rotating quarter wave plate (RQWP). The integrated polarimeter optical spectrum
analyzer (POSA) allows determining both SOP and degree of polarization (DOP)
with high spectral resolution.

The SOP can be described as a point in the Poincar\'{e} unit sphere [see
Fig. \ref{FigSOP}(a)]. The colinear vertical, horizontal, diagonal and
anti-diagonal SOP are denoted by $\left\vert V\right\rangle $, $\left\vert
H\right\rangle $, $\left\vert D\right\rangle =2^{-1/2}\left( \left\vert
H\right\rangle +\left\vert V\right\rangle \right) $ and $\left\vert
A\right\rangle =2^{-1/2}\left( \left\vert H\right\rangle -\left\vert
V\right\rangle \right) $ , respectively, whereas the circular right-hand and
left-hand SOP are denoted by $\left\vert R\right\rangle =2^{-1/2}\left(
\left\vert H\right\rangle -i\left\vert V\right\rangle \right) $ and $%
\left\vert L\right\rangle =2^{-1/2}\left( \left\vert H\right\rangle
+i\left\vert V\right\rangle \right) $, respectively. The unit vectors in the
Poincar\'{e} sphere corresponding to the SOP $\left\vert V\right\rangle $, $%
\left\vert H\right\rangle $, $\left\vert D\right\rangle $, $\left\vert
A\right\rangle $, $\left\vert R\right\rangle $ and $\left\vert
L\right\rangle $, are $\mathbf{\hat{x}}_{3}$, $-\mathbf{\hat{x}}_{3}$, $%
\mathbf{\hat{x}}_{1}$, $-\mathbf{\hat{x}}_{1}$, $-\mathbf{\hat{x}}_{2}$ and $%
\mathbf{\hat{x}}_{2}$, respectively.

In this work POSA is employed for studying two systems. The first is a
ferrimagnetic sphere resonator (FMSR) made of
yttrium iron garnet (YIG) (see sections \ref{SecFSR_M} and \ref{SecFSR_TDT}%
). It has been recently demonstrated that optical single sideband modulation
(SSM) can be implemented in the telecom band using a FMSR \cite{Haigh_133602,Chai_820}. Section \ref{SecUSOC} is devoted
to POSA measurements of a cryogenic fiber loop laser, which is operated in a
region where an unequally spaced optical comb (USOC) is formed \cite%
{Buks_128591}. For both systems, theoretical interpretation of POSA
measurement results is discussed.

\section{POSA}

\label{SecPOSA}

\begin{figure}[tbp]
\begin{center}
\includegraphics[width=3.2in,keepaspectratio]{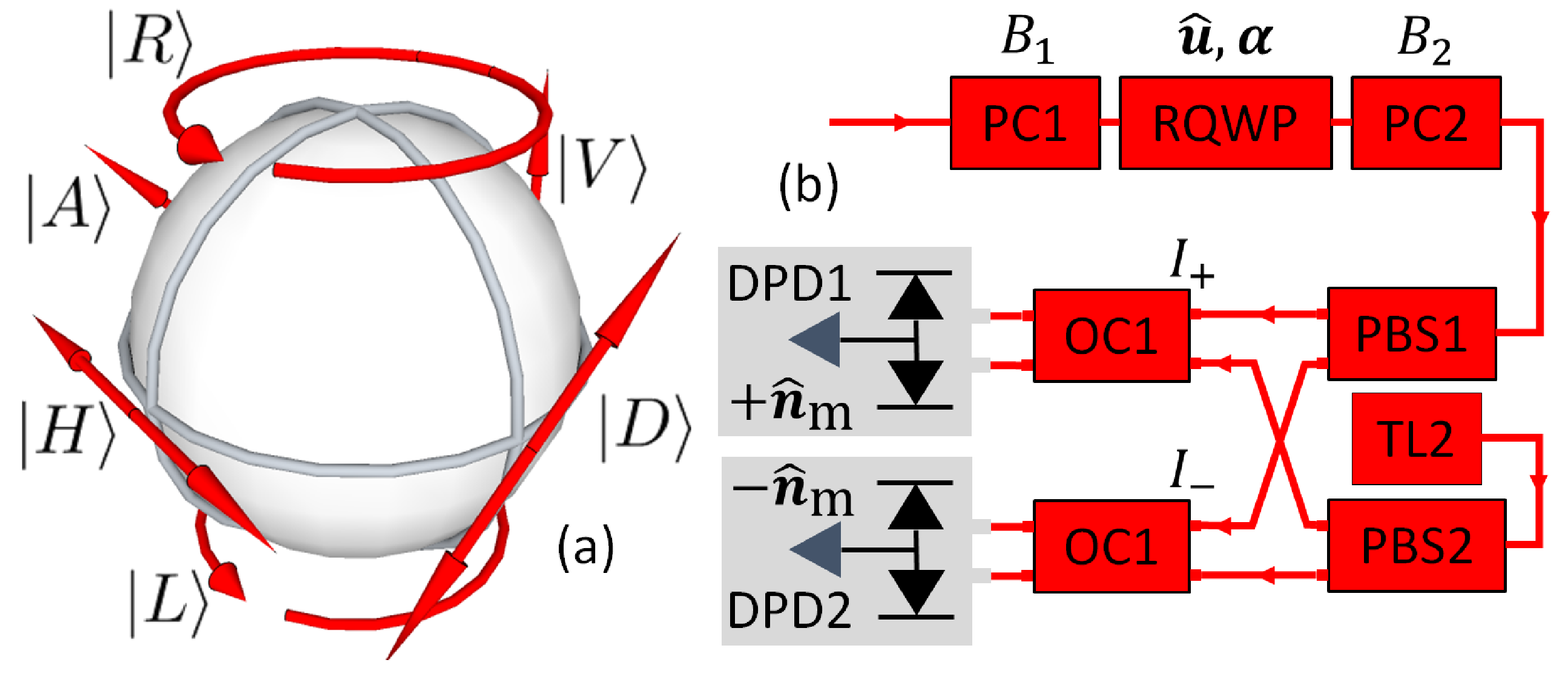}
\end{center}
\caption{{}SOP. (a) The Poincar\'{e} sphere. (b) POSA setup.}
\label{FigSOP}
\end{figure}

POSA setup is schematically shown in Fig. \ref{FigSOP}(b). The input section
contains two polarization controllers [PC1 and PC2 in Fig. \ref{FigSOP}(b)],
and a RQWP (which is based on a rotating stage placed between
two fiber collimators). Heterodyne detection in the coherent
OSA section is performed using a wavelength tunable laser [TL2 in Fig. \ref%
{FigSOP}(b)], two polarized beam splitters [PBS1 and PBS2 in Fig. \ref%
{FigSOP}(b)], two 50:50 optical couplers [OC1 and OC2 in Fig. \ref{FigSOP}%
(b)], and two differential photodetectors [DPD1 and DPD2 in Fig. \ref{FigSOP}%
(b)].

In some other RQWP-based polarimeters, the RQWP is directly
attached to a PBS and a photodetector \cite{goldstein2017polarized}. In
contrast, a single mode fiber is used in our setup to connect the RQWP to
the coherent OSA section [see Fig. \ref{FigSOP}(b)]. This inter-section
connection gives rise to a unitary transformation denoted by $B_{2}$. The
transformation $B_{2}$ can be manipulated using PC2, however, it is a priori
unknown. Consequently, the extraction process of SOP from POSA data, which
is explained below, is more complicated than the process that is commonly
employed in other RQWP-based polarimeters, for which $B_{2}$ represents the
identity transformation \cite{goldstein2017polarized}.

Let $\rho $ be the density matrix of a given SOP, and let $P_{\mathrm{m}}$
be the matrix representation of a given projection operator (associated with
a given polarization filter). The $2\times 2$ matrices $\rho $ and $P_{%
\mathrm{m}}$ are expressed as $\rho =\left( 1/2\right) \left( 1+\gamma 
\mathbf{\hat{n}}\cdot \boldsymbol{\sigma }\right) $ and $P_{\mathrm{m}%
}=\left( 1/2\right) \left( 1+\mathbf{\hat{n}}_{\mathrm{m}}\cdot %
\boldsymbol{\sigma }\right) $, where both $\mathbf{\hat{n}}=\gamma
^{-1}\left( P_{1},P_{2},P_{3}\right) =\left( \sin \theta _{\mathrm{n}}\cos
\varphi _{\mathrm{n}},\cos \theta _{\mathrm{\ n}},\sin \theta _{\mathrm{n}%
}\sin \varphi _{\mathrm{n}}\right) $ and $\mathbf{\ \hat{n}}_{\mathrm{m}%
}=\left( m_{1},m_{2},m_{3}\right) =\left( \sin \theta \cos \varphi ,\cos
\theta ,\sin \theta \sin \varphi \right) $ are real unit vectors 
(the over-hat symbol is used to denote unit vectors), $0\leq
\gamma \leq 1$ is the DOP, and $\boldsymbol{\sigma }=\left( \sigma
_{1},\sigma _{2},\sigma _{3}\right) $ is the Pauli matrix vector%
\begin{equation}
\sigma _{1}=\left( 
\begin{array}{cc}
0 & 1 \\ 
1 & 0%
\end{array}%
\right) ,\;\sigma _{2}=\left( 
\begin{array}{cc}
0 & -i \\ 
i & 0%
\end{array}%
\right) ,\;\sigma _{3}=\left( 
\begin{array}{cc}
1 & 0 \\ 
0 & -1%
\end{array}%
\right) \;.  \label{Pauli}
\end{equation}%
Using the relation $\left( \boldsymbol{\sigma }\cdot \mathbf{a}\right)
\left( \boldsymbol{\sigma }\cdot \mathbf{b}\right) =\mathbf{a}\cdot \mathbf{b%
}+i\boldsymbol{\sigma }\cdot \left( \mathbf{a}\times \mathbf{b}\right)$, one
finds that the probability $p$ to find the SOP pointing in the $\mathbf{\hat{%
n}}_{\mathrm{m}}$ direction is given by%
\begin{equation}
p=\func{Tr}\left( \rho P_{\mathrm{m}}\right) =\frac{1+\gamma \mathbf{\hat{n}}%
\cdot \mathbf{\hat{n}}_{\mathrm{m}}}{2}\;.  \label{p SOP}
\end{equation}

While the unit vector $\mathbf{\hat{n}}_{\mathrm{m}}$ is associated with the
polarization filter of OSA trace 1 (DPD1), OSA trace 2 (DPD2) represents the
orthogonal SOP corresponding to the unit vector $-\mathbf{\hat{n}}_{\mathrm{m%
}}$. For a given input signal having intensity $I$, the OSA trace 1 (2)
signal intensity $I_{+}$ ($I_{-}$) is given by $I_{\pm }/I=\left( 1/2\right)
\left( 1\pm \gamma \mathbf{\hat{n}}\cdot \mathbf{\hat{n}}_{\mathrm{m}%
}\right) $ [see Eq. (\ref{p SOP})], and thus the following holds%
\begin{equation}
\gamma \mathbf{\hat{n}}\cdot \mathbf{\hat{n}}_{\mathrm{m}}=\frac{I_{+}-I_{-}%
}{I_{+}+I_{-}}\;.  \label{gamma*n*n_m}
\end{equation}

Any loss-less linear SOP transformation can be described using a unitary
Jones matrix $B\left( \mathbf{\hat{u}},\phi \right) $ given by \cite%
{Potton_717}%
\begin{equation}
B\left( \mathbf{\hat{u}},\phi \right) \dot{=}\exp \left( -\frac{i%
\boldsymbol{\ \sigma }\cdot \mathbf{\hat{u}}\phi }{2}\right) =\mathbf{1}\cos 
\frac{\phi }{2}-i\boldsymbol{\sigma }\cdot \mathbf{\hat{u}}\sin \frac{\phi }{%
2}\;,  \label{B(u,phi)}
\end{equation}%
where $\mathbf{\hat{u}}$ is a unit vector and $\phi $ is a rotation angle.
By expressing the unit vector $\mathbf{\hat{n}}$ as $\mathbf{\hat{n}}=%
\mathbf{n}_{\parallel }+\mathbf{n}_{\perp }$, where $\mathbf{n}_{\parallel }=%
\mathbf{\ \left( \hat{u}\cdot \mathbf{\hat{n}}\right) \hat{u}}$ (parallel
component of $\mathbf{\hat{n}}$ in the $\mathbf{\hat{u}}$ direction) and $%
\mathbf{n}_{\perp }=\mathbf{\hat{u}}\times \left( \mathbf{\hat{n}}\times 
\mathbf{\hat{u}}\right) $ (perpendicular component), one finds that%
\begin{equation}
B^{{}}\left( \mathbf{\hat{u}},\phi \right) \left( \boldsymbol{\sigma }\cdot 
\mathbf{\hat{n}} \right) B^{\dag }\left( \mathbf{\hat{u}},\phi \right) =%
\boldsymbol{\ \sigma }\cdot \mathbf{\hat{n}}_{\mathrm{T}}\;,  \label{n2nT}
\end{equation}%
where%
\begin{equation}
\mathbf{\hat{n}}_{\mathrm{T}}=\mathbf{n}_{\parallel }+\mathbf{n}_{\perp
}\cos \phi +\left( \mathbf{\hat{u}}\times \mathbf{n}_{\perp }\right) \sin
\phi \;.  \label{nT}
\end{equation}%
The transformation from $\mathbf{\hat{n}}$ to $\mathbf{\hat{n}}_{\mathrm{T}}$
(\ref{nT}) is a rotation about the $\mathbf{\hat{u}}$ axis with angle $\phi $%
.

The unit vector $\mathbf{\hat{u}}$ corresponding to the RQWP is given by $%
\mathbf{\hat{u}}=-\sin \left( 2\alpha \right) \mathbf{\hat{x}}_{1}+\cos
\left( 2\alpha \right) \mathbf{\hat{x}}_{3}$, where $\alpha $ is the RQWP
axis angle, and the rotation angle $\phi $ is given by $\phi =\pi /2$, thus
[see Eqs. (\ref{n2nT}) and (\ref{nT})]%
\begin{align}
\gamma \mathbf{\hat{n}}_{\mathrm{T}}\cdot \mathbf{\hat{n}}_{\mathrm{m}}&
=\gamma \left( \mathbf{\left( \hat{u}\cdot \mathbf{\hat{n}}\right) \hat{u}}%
+\left( \mathbf{\hat{u}}\times \mathbf{n}\right) \right) \cdot \mathbf{\hat{n%
}}_{\mathrm{m}}  \notag \\
& =a_{0}+a_{1}\cos \left( 2\alpha \right) +b_{1}\sin \left( 2\alpha \right) 
\notag \\
& +a_{2}\cos \left( 4\alpha \right) +b_{2}\sin \left( 4\alpha \right) \;, 
\notag \\
&  \label{gamma*n_T*n_m}
\end{align}%
where%
\begin{align}
a_{0}& =\frac{P_{1}m_{1}+P_{3}m_{3}}{2}\;,  \label{a0 V1} \\
a_{1}& =P_{1}m_{2}-P_{2}m_{1}\;,  \label{a1 V1} \\
b_{1}& =P_{3}m_{2}-P_{2}m_{3}\;,  \label{b1 V1} \\
a_{2}& =\frac{P_{3}m_{3}-P_{1}m_{1}}{2}\;,  \label{a2 V1} \\
b_{2}& =-\frac{P_{1}m_{3}+P_{3}m_{1}}{2}\;.  \label{b2 V1}
\end{align}%
The following holds [see Eqs. (\ref{a2 V1}) and (\ref{b2 V1})]%
\begin{equation}
a_{2}^{2}+b_{2}^{2}=\frac{\gamma ^{2}\sin ^{2}\theta \sin ^{2}\theta _{%
\mathrm{n}}}{4}\;.  \label{a_2^2+b2^2 V1}
\end{equation}

The Poincar\'{e} vector $\mathbf{P}=\left( P_{1},P_{2},P_{3}\right) $ can be
extracted from Eqs. (\ref{a0 V1})-(\ref{b2 V1}), and the measured values of $%
a_{0}$, $a_{1}$, $b_{1}$, $a_{2}$ and $b_{2}$, provided that the
transformation $B_{2}$, which determines the unit vector $\mathbf{\ \hat{n}}%
_{\mathrm{m}}=\left( m_{1},m_{2},m_{3}\right) $, is given. POSA calibration
(which is needed because $B_{2}$ is a priori unknown) is done by varying the
angles $\theta _{\mathrm{n}}$ and $\varphi _{\mathrm{n}}$ associated with
the SOP transformation $B_{1}$ controlled by PC1, while keeping $\theta $
and $\varphi $ unchanged ($\theta $ and $\varphi $ are determined by the SOP
transformation $B_{2}$ from the RQWP to the coherent OSA section, which can
be manipulated using PC2). This process allows determining the
angle $\theta _{\mathrm{n}}$ using the relation [see Eq. (\ref{a_2^2+b2^2 V1}%
)]%
\begin{equation}
\sin ^{2}\theta _{\mathrm{n}}=\frac{a_{2}^{2}+b_{2}^{2}}{\max_{\theta _{%
\mathrm{n}},\varphi _{\mathrm{n}}}\left( a_{2}^{2}+b_{2}^{2}\right) }\;.
\end{equation}

The unit vector $\mathbf{\hat{n}}_{\mathrm{m}}=\left( \sin \theta \cos
\varphi ,\cos \theta ,\sin \theta \sin \varphi \right) $ can be manipulated
by PC2 [see Fig. \ref{FigSOP}(b)]. As can be seen from Eq. (\ref{a_2^2+b2^2
V1}), the angle $\theta $ can be tuned such that $\sin ^{2}\theta =1$, by
maximizing $a_{2}^{2}+b_{2}^{2}$ using PC2. For this case $%
a_{1}^{2}+b_{1}^{2}+4\left( a_{2}^{2}+b_{2}^{2}\right) =\gamma ^{2}$. This
PC2 calibration, which greatly simplifies SOP extraction from POSA data, has
been performed prior to all measurements with both a driven FMSR (sections \ref{SecFSR_M} and \ref{SecFSR_TDT}) and with a
cryogenic fiber loop laser (section \ref{SecUSOC}).

\section{FMSR modulator}

\label{SecFSR_M}

FMSR are widely employed as magnetically tunable microwave filters having high quality factor. The FMSR ellipsoidal shape allows uniform magnetization. Magneto-optical (MO) coupling \cite%
{Rameshti_1,Kusminskiy_299,Zhu_2012_11119,Juraschek_094407,Bittencourt_014409,Zhang_123605,VanderZiel_190,Zheng_2303_16313,Stancil_Spin}
between FMSR optical \cite%
{Oraevsky_377,Schunk_30795,Gorodetsky_33} and magnetic Walker \cite%
{Walker_390} modes can be used for implementing optical modulation \cite%
{Haigh_133602,Osada_103018,Osada_223601,Sharma_094412,Sharma_087205,Almpanis_184406,Zivieri_165406,Desormiere_379,Liu_3698,Chai_820,Zhu_1291,Li_040344,Nayak_193905}%
. The sequence of sidebands that are generated in a FMSR-based modulator can be determined from Brillouin scattering
selection rules \cite%
{Haigh_133602,Osada_103018,Osada_223601,Sharma_094412,Almpanis_184406,Zivieri_165406,Desormiere_379,Liu_5452,Sandercock_1729,Hu_307,Ghasemian_12757,Kusminskiy_033821}%
, and angular momentum conservation in photon-magnon scattering \cite%
{Wettling_211,cottam1986light,Liu_060405,Haigh_143601,Hisatomi_207401,Hisatomi_174427,Wu_023711}%
. Contrary to some other modulation techniques (such as amplitude, phase and
frequency modulations) \cite{lazzarini2008asymmetric,Li_1,Shimotsu_364},
symmetry between Stokes and anti-Stokes sidebands can be broken in %
FMSR-based modulation \cite%
{Haigh_133602,Chai_820,Nayak_193905}. In particular, the method of single
sideband modulation (SSM), which allows reducing both transmission power and
bandwidth, can be implemented using FMSR modulation.

\begin{figure}[tbp]
\begin{center}
\includegraphics[width=3.2in,keepaspectratio]{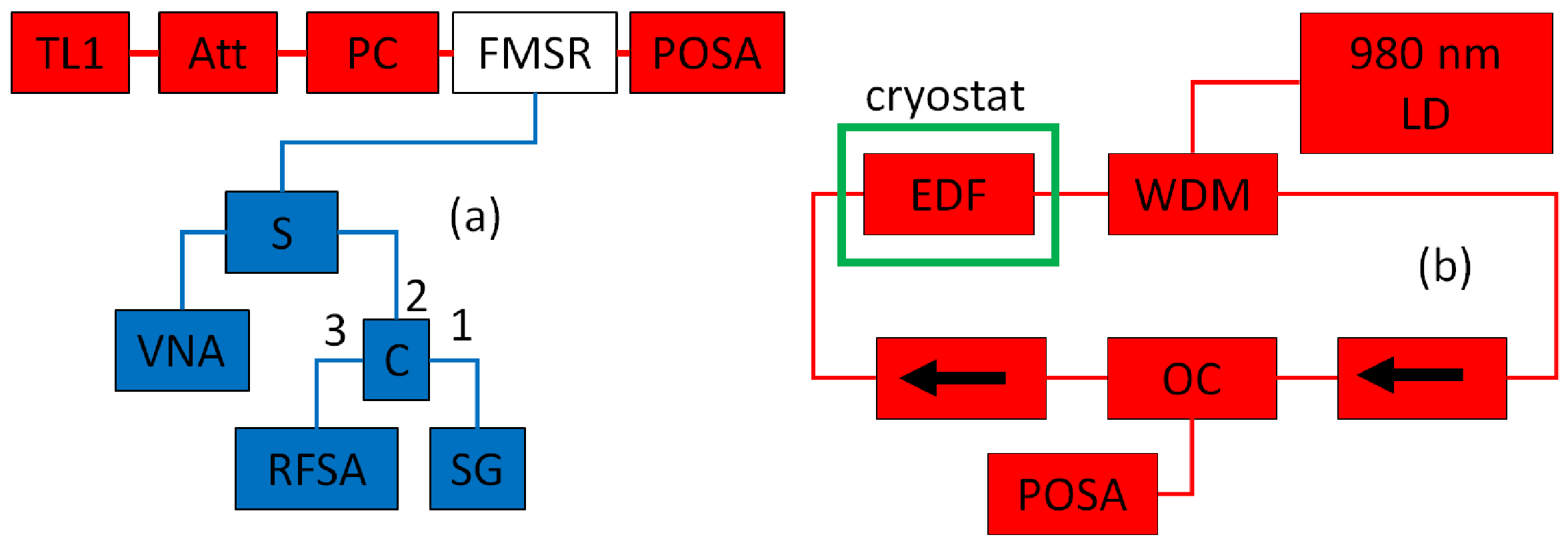} 
\end{center}
\caption{{}Experimental setup. Optical components and fibers are red
colored, whereas blue color is used to label MW components and coaxial
cables. (a) Optical fibers are installed on both sides of the FMSR for transmission of light through the sphere (see Ref. \cite{Nayak_193905} for more details on the experimental setup). (b) The
cryogenic fiber loop laser (see Ref. \cite{Buks_2951} for more details). }
\label{FigSetup}
\end{figure}

The FMSR modulator is schematically shown in Fig.~%
\ref{FigSetup}(a). An FMSR made of YIG having
radius of $R_{\mathrm{s}}=125\func{%
\mu%
m}$ is held by two ceramic ferrules (CF). The two CFs, which are held by a
concentric sleeve, provide transverse alignment for both input and output
single mode optical fibers. All optical measurements are performed in the
telecom band, in which YIG has refractive index of $2.19$ and absorption
coefficient of $\left( 0.5\unit{m}\right) ^{-1}$ \cite{Onbasli_1}. The
intensity and SOP of light illuminating the FMSR are controlled by an optical attenuator (Att) and a PC, respectively. The blue-colored microwave  (MW) components shown in Fig. \ref{FigSetup}(a) allow both driving and detection of FMSR magnetic resonances. The FMSR is inductively coupled to a microwave loop
antenna (MWA). Magnetic resonances are identified using a vector network analyzer
(VNA). A signal generator (SG) is employed for driving, and the response is monitored using a radio frequency spectrum analyzer (RFSA). A circulator (C) and  a splitter (S) are employed to direct the input and output microwave signals [see Fig.~\ref{FigSetup}(a)].

The angular frequency of the FMSR Kittel mode $%
\omega _{\mathrm{m}}$ is approximately given by $\omega _{\mathrm{m}}=\mu
_{0}\gamma _{\mathrm{e}}H_{\mathrm{s}}$, where $H_{\mathrm{s}}$ is the
static magnetic field, $\mu _{0}$ is the free space permeability, and $%
\gamma _{\mathrm{e}}/2\pi =28\unit{GHz}\unit{T}^{-1}$\ is the gyromagnetic
ratio \cite{Fletcher_687,sharma2019cavity,Stancil_Spin,Jin_thesis}. The
applied static magnetic field $\mathbf{H}_{\mathrm{s}}$ is controlled by
adjusting the relative position of a magnetized Neodymium using a motorized
stage. The static magnetic field $\mathbf{H}_{\mathrm{s}}$ is normal to the
unit vector $\mathbf{\hat{k}}$ pointing in the light propagation direction,
and the MWA driving magnetic field is nearly parallel to $\mathbf{\hat{k}}$.
The FMSR is installed such that its
crystallographic direction $\left[ 111\right] $ is aligned to $\mathbf{\hat{k%
}}$.

\begin{figure}[tbp]
\begin{center}
\includegraphics[width=3.2in,keepaspectratio]{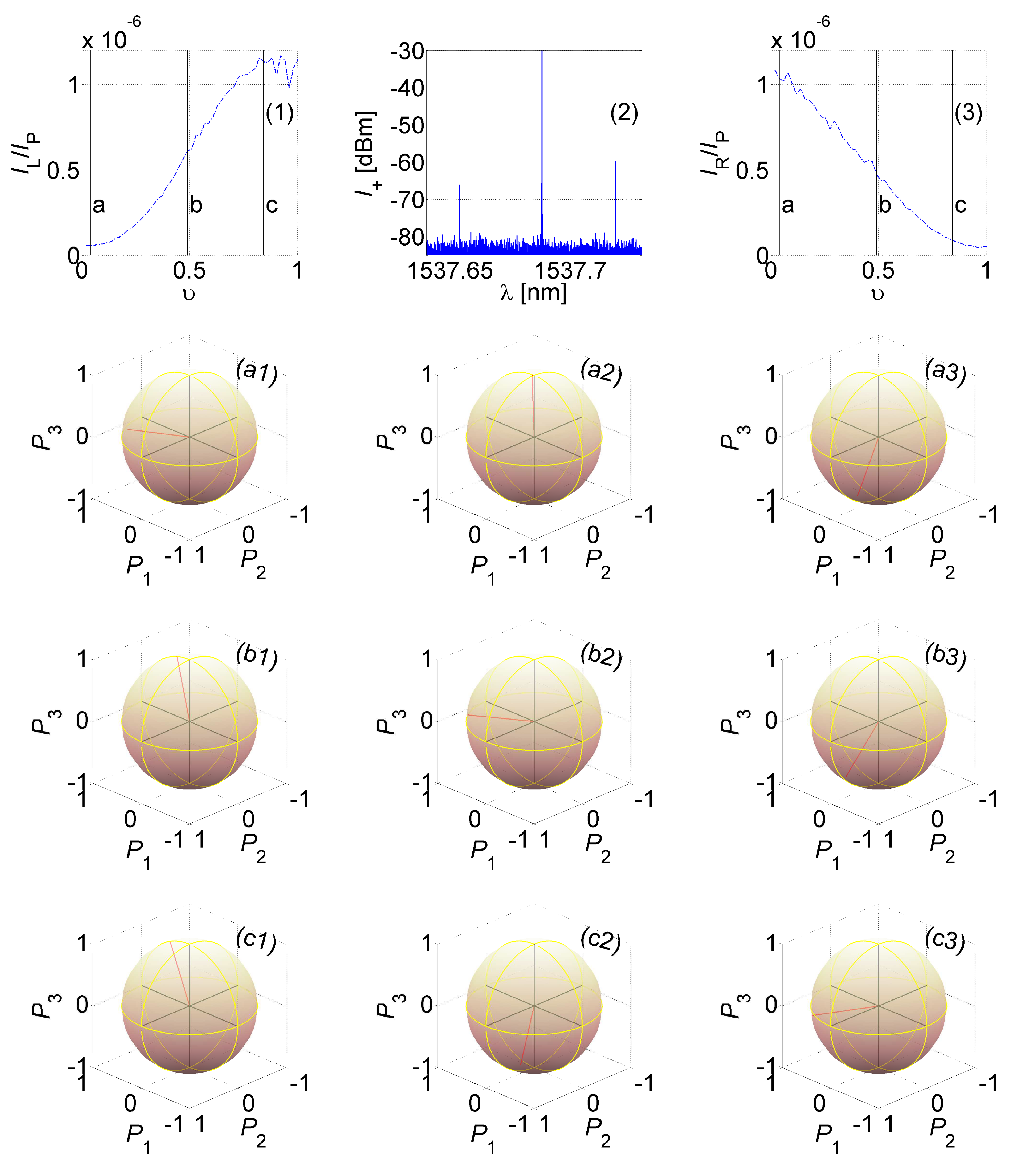}
\end{center}
\caption{{}FMSR. Normalized sideband intensities $%
I_{\mathrm{L}}/I_{\mathrm{P}}$ and $I_{\mathrm{R}}/I_{\mathrm{P}}$ are shown
in (1) and (3), respectively ($I_{\mathrm{P}}$ is the central peak
intensity). The letters a, b and c in the Poincar\'{e} plots' labels refer
to values of $\protect\upsilon $ indicated in (1) and (3). The numbers 1, 2
and 3 in the Poincar\'{e} plots' labels refer to the left sideband, central
peak and right sideband, respectively. The Poincar\'{e} vectors
are red colored.}
\label{FigFSR}
\end{figure}

The plot in Fig. \ref{FigFSR}(2) shows POSA DPD1 signal intensity [see Fig. %
\ref{FigSOP}(b)] as a function of TL2 optical wavelength $\lambda$. For this
plot the TL1 [see Fig.~\ref{FigSetup}(a)] power is $P_{\mathrm{L}}=6\func{mW}
$ and wavelength is $\lambda_{\mathrm{L}}=1537.7\func{nm}$. The MWA driving, which
has power of $12\func{dBm}$, is frequency tuned to resonance at $\omega _{%
\mathrm{m}}/\left(2\pi\right)=4.1755\func{GHz}$. The driving-induced
sidebands around the central peak [see Fig. \ref{FigFSR}(2)] have
wavelengths $\lambda_{\mathrm{L}}\left(1\pm \lambda_{\mathrm{L}}\omega _{%
\mathrm{m}}/\left(2\pi c\right)\right)=\lambda_{\mathrm{L}}\pm32.9\func{pm}$%
, where $c$ is the vacuum speed of light. Sidebands' intensities strongly
depend on the input light SOP. This dependency is explored in the next
section, which is devoted to the FMSR transverse
dielectric tensor.

\section{FMSR transverse dielectric tensor}

\label{SecFSR_TDT}

In a Cartesian coordinate system, for which the $[100]$, $[010]$, and $[001]$
crystallographic directions are pointing in the $x$, $y$ and $z$ axes,
respectively, the dielectric tensor is represented by a $3\times 3$ matrix
denoted by $\varepsilon $ \cite{Freiser_152,Boardman_197,Boardman_388}. The
Onsager reciprocal relations, which originate from time-reversal symmetry,
read $\epsilon _{nm}\left( \mathbf{M}\right) =\epsilon _{mn}\left( -\mathbf{M%
}\right) $, where $\mathbf{M}=\left( M_{1},M_{2},M_{3}\right) \mathbf{\ }$is
the magnetization vector, and $n,m\in \left\{ 1,2,3\right\} $. The MO
Stoner--Wohlfarth energy density is given by $u=\left( 1/4\right) \mathbf{E}%
^{\ast }\varepsilon ^{{}}\mathbf{E}^{\mathrm{T}}$, where $\mathbf{E}=\left(
E_{1},E_{2},E_{3}\right) $ is the electric field vector phasor.

To second order in the magnetization $\mathbf{M}$, the dielectric tensor can
be expressed as $\varepsilon =\varepsilon _{\mathrm{r}}+\varepsilon
_{1}+\varepsilon _{2} $, where $\varepsilon _{\mathrm{r}}$ is the relative
permittivity. The first order (in $\mathbf{M}$) contribution to the
dielectric tensor is given by $\varepsilon _{1}=iQ_{\mathrm{s}}\epsilon
_{ijk}M_{k}$, where $\epsilon _{ijk}$ is the Levi-Civita symbol. For YIG in
saturated magnetization $Q_{\mathrm{s}}\simeq 10^{-4}$ \cite{Wood_1038}.
Crystal cubic symmetry implies that the second order contribution $%
\varepsilon _{2}$ can be expressed as $\varepsilon _{2}=g\func{diag}\left(
M_{1}^{2},M_{2}^{2},M_{3}^{2}\right) +g_{12}\mathbf{M}^{{}}\mathbf{M}^{%
\mathrm{T}}+2g_{44}\mathbf{M}^{\mathrm{T}}\mathbf{M}^{{}}$, where $g$, $%
g_{12}$, and $g_{44}$ are constants \cite{Prokhorov_339,prabhakar2009spin}
(in the notation used in Ref. \cite{Prokhorov_339} $g=g_{11}-g_{12}-2g_{44}$%
).

Consider the coordinate transformation $\mathbf{M}^{\prime }=\mathbf{M}^{{}}%
\mathcal{R}^{{}}$, where $\mathcal{R}$ is a $3\times 3$ rotation matrix. The
energy density can be expressed as $u=\left( 1/4\right) \mathbf{E}^{\prime
\ast }\varepsilon ^{\prime }\mathbf{E}^{\prime \mathrm{T}}$, where $\mathbf{E%
}^{\prime }=\mathbf{E}^{{}}\mathcal{R}^{{}}$ is the transformed value of the
electric field $\mathbf{E}^{{}}$, and where $\varepsilon ^{\prime }=\mathcal{%
R}^{-1}\varepsilon ^{{}}\mathcal{R}^{{}}$ is the transformed value of the
dielectric tensor $\varepsilon ^{{}}$ (note that $\mathcal{R}^{\mathrm{T}}=%
\mathcal{R}^{-1}$ and that $\mathcal{R}^{{}}$ is real). The corresponding
transformed first (second) order contribution to the dielectric tensor is
denoted by $\varepsilon _{1}^{\prime }$ ($\varepsilon _{2}^{\prime }$).

The following holds $\mathbf{E}^{\ast }\varepsilon _{1}^{{}}\mathbf{E}^{%
\mathrm{T}}=-iQ_{\mathrm{s}}\mathbf{E}^{\ast }\cdot \left( \mathbf{M}\times 
\mathbf{E}\right) $, thus using the matrix identity $\left( \mathcal{R}%
\mathbf{X}\right) \times \left( \mathcal{R}\mathbf{Y}\right) =$ $\mathcal{R}%
\left( \mathbf{X}\times \mathbf{Y}\right) $, where $\mathbf{X}$ and $\mathbf{%
Y}$ are vectors and $\mathcal{R}$ is a rotation matrix, one finds that $%
\varepsilon _{1}^{\prime }\left( \mathbf{M}^{\prime }\right) =\varepsilon
_{1}^{{}}\left( \mathbf{M}^{{}}\right) $, i.e. $\varepsilon _{1}^{\prime
}=iQ_{\mathrm{s}}\epsilon _{ijk}M_{k}^{\prime }$. The transformed second
order contribution $\varepsilon _{2}^{\prime }$ is given by $\varepsilon
_{2}^{\prime }=gE_{\mathrm{g}}^{\prime }+g_{12}\mathbf{M}^{\prime }\left( 
\mathbf{M}^{\prime }\right) ^{\mathrm{T}}+2g_{44}\left( \mathbf{M}^{\prime
}\right) ^{\mathrm{T}}\mathbf{M}^{\prime }$, where%
\begin{equation}
E_{\mathrm{g}}^{\prime }=\sum_{n=1}^{3}\mathcal{R}^{-1}\eta _{n,n}\mathcal{R}%
^{{}}\left( \mathbf{M}^{\prime }\right) ^{\mathrm{T}}\mathbf{M}^{\prime }%
\mathcal{R}^{-1}\eta _{n,n}\mathcal{R}^{{}}\;,  \label{E_g prime}
\end{equation}%
and where $\eta _{n,m}$ is a $3\times 3$ matrix, whose entries are given by $%
\left( \eta _{n,m}\right) _{n^{\prime },m^{\prime }}=\delta
_{n^{{}},n^{\prime }}\delta _{m^{{}},m^{\prime }}$. Note that only the term $%
E_{\mathrm{g}}^{\prime }$ given by Eq. (\ref{E_g prime}) gives rise to
dependency of $\varepsilon _{2}^{\prime }$ on crystallographic directions.

The rotation matrix $\mathcal{R}$, which is given by%
\begin{equation}
\mathcal{R}=\left( 
\begin{array}{ccc}
\frac{\cos \phi }{\sqrt{2}}-\frac{\sin \phi }{\sqrt{6}} & -\frac{\cos \phi }{%
\sqrt{6}}-\frac{\sin \phi }{\sqrt{2}} & \frac{1}{\sqrt{3}} \\ 
\frac{2\sin \phi }{\sqrt{6}} & \frac{2\cos \phi }{\sqrt{6}} & \frac{1}{\sqrt{%
3}} \\ 
-\frac{\cos \phi }{\sqrt{2}}-\frac{\sin \phi }{\sqrt{6}} & -\frac{\cos \phi 
}{\sqrt{6}}+\frac{\sin \phi }{\sqrt{2}} & \frac{1}{\sqrt{3}}%
\end{array}%
\right) \;,
\end{equation}%
maps the $z$ direction to $\left[ 111\right] $ (which is parallel to light
propagation direction $\mathbf{\hat{k}}$), and the $xy$ plane to $\left(
111\right) $, with a variable angle $\phi $. Consider the case where the
static magnetic field is applied parallel to the $x^{\prime }$ direction. It
is assumed that $M_{1}^{\prime }\simeq M_{0}$, $M_{2}^{\prime }/M_{0}\equiv
\mu _{2}\ll 1$ and $M_{3}^{\prime }/M_{0}\equiv \mu _{3}\ll 1$, where $M_{0}$
is the saturation magnetization.

In the truncation approximation, the transverse dielectric tensor is taken
to be the $2\times 2$ upper left block of the $3\times 3$ tensor $%
\varepsilon ^{\prime }$. Moreover, terms independent on $Q_{\mathrm{s}}$, $%
\mu _{2}$ and $\mu _{3}$ are disregarded. In this approximation $\varepsilon
^{\prime }-\varepsilon _{\mathrm{r}}$ is represented by a $2\times 2$ matrix
given by $M_{0}^{2}\left( \mu _{3}\mathbf{a}^{\prime }+\mu _{2}\mathbf{a}%
^{\prime \prime }\right) \cdot \boldsymbol{\sigma }$, where $\mathbf{a}%
^{\prime } $ ($\mathbf{a}^{\prime \prime }$) is the real (imaginary) part of
the vector $\mathbf{a}=\mathbf{a}^{\prime }+i\mathbf{a}^{\prime \prime }$,
which is given by (to first order in $Q_{\mathrm{s}}$)%
\begin{equation}
\mathbf{a}=-\frac{\sqrt{2}g}{3}\left( \cos 3\phi -\frac{3i\left( \frac{1}{3}+%
\frac{2g_{44}}{g}\right) }{\sqrt{2}},\frac{3Q_{\mathrm{s}}}{\sqrt{2}M_{0}g}%
,\sin 3\phi \right) \;,  \label{k=}
\end{equation}%
and where $\boldsymbol{\sigma }$ is the Pauli matrix vector [see Eq. (\ref%
{Pauli})]. Alternatively 
\begin{equation}
\varepsilon ^{\prime }-\varepsilon _{\mathrm{r}}=\frac{M_{0}^{2}}{2}\left(
\mu _{+}\mathbf{a}^{\ast }+\mu _{-}\mathbf{a}^{{}}\right) \cdot %
\boldsymbol{\sigma }\;,  \label{TDT}
\end{equation}%
where $\mu _{\pm }=\mu _{3}\pm i\mu _{2}$.

The Hermitian matrices $\left( \mathbf{a}^{\prime \prime }\times \mathbf{a}%
^{\prime }\right) \cdot \boldsymbol{\sigma }$, $S^{\dag }S^{{}}$ and $%
S^{{}}S^{\dag }$, where $S=\mathbf{a}\cdot \boldsymbol{\sigma }$ [see Eq. (%
\ref{TDT})], share the same eigenvectors, which are denoted by $\left\vert 
\mathbf{a}_{+}\right\rangle $ and $\left\vert \mathbf{a}_{-}\right\rangle $.
The transformed matrix $\tilde{S}=U^{-1}SU^{{}}$, where $U$ is the $2\times 2
$ unitary transformation matrix corresponding to the basis $\left\{
\left\vert \mathbf{a}_{+}\right\rangle ,\left\vert \mathbf{a}%
_{-}\right\rangle \right\} $, has a hollow form [see Eq. (6.150) of Ref. 
\cite{Buks_QMLN}]. The matrix representation of $\varepsilon ^{\prime
}-\varepsilon _{\mathrm{r}}$ in that basis, which is denoted by $\varepsilon
_{\mathrm{h}}$, is given by%
\begin{equation}
\varepsilon _{\mathrm{h}}=\frac{M_{0}^{2}}{2}\left( 
\begin{array}{cc}
0 & \mu _{-}\zeta _{1}^{{}}+\mu _{+}\zeta _{2}^{\ast } \\ 
\mu _{-}\zeta _{2}^{{}}+\mu _{+}\zeta _{1}^{\ast } & 0%
\end{array}%
\right) \;,  \label{epsilon_h}
\end{equation}%
where $\zeta _{1}\zeta _{2}=\mathbf{a}\cdot \mathbf{a}$, $\left\vert \zeta
_{1}\right\vert ^{2}+\left\vert \zeta _{2}\right\vert ^{2}=2\mathbf{a}%
^{{}}\cdot \mathbf{a}^{\ast }$\ and $\left\vert \zeta _{2}\right\vert
^{2}-\left\vert \zeta _{1}\right\vert ^{2}=4\left\vert \mathbf{a}^{\prime
\prime }\times \mathbf{a}^{\prime }\right\vert $.

Both scattering selection rules and sideband asymmetry (between Stokes and
anti-Stokes) can be characterized in terms of the ratio $\zeta _{\mathbf{a}}$%
, which is defined by [see Eq. (\ref{epsilon_h}), and note that $\left\vert
\zeta _{\mathbf{a}}\right\vert \leq 1$]%
\begin{equation}
\zeta _{\mathbf{a}}=\frac{\left\vert \zeta _{2}\right\vert ^{2}-\left\vert
\zeta _{1}\right\vert ^{2}}{\left\vert \zeta _{1}\right\vert ^{2}+\left\vert
\zeta _{2}\right\vert ^{2}}=\frac{2\left\vert \mathbf{a}^{\prime \prime
}\times \mathbf{a}^{\prime }\right\vert }{\left\vert \mathbf{a}^{\prime
}\right\vert ^{2}+\left\vert \mathbf{a}^{\prime \prime }\right\vert ^{2}}\;.
\label{zeta_a}
\end{equation}%
Note that sideband asymmetry is possible only when $\left\vert \mathbf{a}%
^{\prime \prime }\times \mathbf{a}^{\prime }\right\vert >0$ [see Eqs. (\ref%
{epsilon_h}) and (\ref{zeta_a})]. For radio and optical communication, it is
well known that sideband asymmetry is impossible when only the amplitude, or
only the frequency, or only the phase, is modulated by a monochromatic
signal. Moreover,with simultaneously applied two different modulation
methods, sideband asymmetry becomes possible only when one modulation method
is phase-shifted with respect to the other. These properties are all
reflected by the condition $\left\vert \mathbf{a}^{\prime \prime }\times 
\mathbf{a}^{\prime }\right\vert >0$.

For YIG $\left( Q_{\mathrm{s}}/\left( M_{0}g\right) \right) ^{2}=44.2$ and $%
2g_{44}/g=-4.0$\ \cite{Stancil_Spin}, hence $\zeta _{\mathbf{a}}\simeq 0.84$
[see Eqs. (\ref{k=}) and (\ref{zeta_a}), and note that the term $\sin
^{2}3\phi \ $has a relatively small effect]. The largest sideband asymmetry
is obtained for input states $\left\vert \mathbf{a}_{+}\right\rangle $ and $%
\left\vert \mathbf{a}_{-}\right\rangle $ [eigenvectors of $\left( \mathbf{a}%
^{\prime \prime }\times \mathbf{a}^{\prime }\right) \cdot \boldsymbol{\sigma
}$]. As can be seen from Eq. (\ref{k=}), for YIG the vector $\mathbf{a}%
^{\prime \prime }\times \mathbf{a}^{\prime } $ is nearly parallel to the
unit vector $\mathbf{\hat{z}}$.

Intensities of both left (shorter wavelength, anti-Stokes) and right (longer
wavelength, Stokes) sidebands can be manipulated using the PC shown in Fig. %
\ref{FigSetup}(a), which controls the input SOP. The vector of three
voltages controlling this PC are denoted by ${\mathbf{V}}_{\mathrm{PC}%
}=\left( V_{\mathrm{PC},1},V_{\mathrm{PC},2},V_{\mathrm{PC},3}\right) $. A
searching procedure is implemented to determine the value ${\mathbf{V}}_{%
\mathrm{PC,L}}$ (${\mathbf{V}}_{\mathrm{PC,R}}$), for which the intensity of
the left (right) sideband, which is denoted by $I_{\mathrm{L}}$ ($I_{\mathrm{%
R}}$), is maximized. The measurements shown in Fig. \ref{FigFSR} are
performed with values of ${\mathbf{V}}_{\mathrm{PC}}$ given by ${\mathbf{V}}%
_{\mathrm{PC}}=\upsilon {\mathbf{V}}_{\mathrm{PC,L}}+\left( 1-\upsilon
\right) {\mathbf{V}}_{\mathrm{PC,R}}$, where $0\leq \upsilon \leq 1$. The
Poincar\'{e} vector $\mathbf{P}=\left( P_{1},P_{2},P_{3}\right) $ can be
extracted from POSA data using the method above-discussed in section \ref%
{SecPOSA}. Poincar\'{e} vectors $\mathbf{P}$ are shown in Fig. \ref{FigFSR}
for three different values of the parameter $\upsilon $, which are indicated
in Fig. \ref{FigFSR}(1) and (3) by the letters a, b and c. The numbers 1, 2
and 3 in the Poincar\'{e} plots' labels in Fig. \ref{FigFSR} refer to the
left sideband, central peak and right sideband, respectively.

A value for the asymmetry parameter given by $\zeta _{\mathbf{a}}=0.92$ is
extracted from the data shown in Fig. \ref{FigFSR}(1) and (3). This result
suggests that the above-discussed theoretical analysis underestimates $\zeta
_{\mathbf{a}}$ (recall that the calculated value is $0.84$) \cite%
{Haigh_143601}. Disagreement in SOP between theory and experiment is
quantified by the parameter $\delta _{\mathrm{P}}=\left\Vert \mathbf{P}_{%
\mathrm{E}}-\mathbf{P}_{\mathrm{T}}\right\Vert /\left\Vert \mathbf{P}_{%
\mathrm{T}}\right\Vert $, where $\mathbf{P}_{\mathrm{E}}$ denotes a measured
Poincar\'{e} vector, and $\mathbf{P}_{\mathrm{T}}$ is theoretically derived
using Eq. (\ref{epsilon_h}). For the data shown in Fig. \ref{FigFSR} $\delta
_{\mathrm{P}}\lesssim 0.13$. The disagreement is partially attributed to
magneto-dichroism (i.e. polarization-dependent absorption), which is
theoretically disregarded. Further study is needed to explore the
effect of magneto-dichroism and other possible mechanisms contributing to
the observed disagreement.

\section{USOC}

\label{SecUSOC}

USOC formation in a cryogenic fiber loop laser has been recently
demonstrated \cite{Buks_128591,Buks_2951}. USOC can be employed as a
tunable multi-mode lasing source \cite{Buks_L051001}. Multimode lasing has a
variety of applications in the fields of sensing, spectroscopy, signal
processing and communication. Multimode lasing in the telecom band has been
demonstrated by integrating Erbium doped fibers (EDF) cooled by liquid
nitrogen into a fiber ring laser \cite%
{Yamashita_1298,Liu_102988,Lopez_085401,Le_3611}. It has been recently
proposed that EDF operating at low temperatures can be used for storing
quantum information \cite%
{Wei_2209_00802,Ortu_035024,Liu_2201_03692,Veissier_195138,Shafiei_F2A}. For
some USOC-based applications, DOP specification is important. As is
discussed below, the POSA high spectral resolution allows such a
specification for individual USOC peaks.

A sketch of the cryogenic fiber loop laser used for studying USOC is shown
in Fig. \ref{FigSetup}(b). The loop is made of an undoped single mode fiber %
(Corning 28), and a $5\func{m}$ long EDF, which is
used for optical amplification. The EDF has absorption of $30$
dB $\unit{ m}^{-1}$ at $1530\unit{nm}$, and mode field diameter of $6.5\mu 
\unit{m}$ at $1550\unit{nm}$. The cold EDF is integrated with
a room temperature fiber loop using a wavelength-division multiplexing (WDM)
device. The EDF is pumped using a $980 \func{nm}$ laser diode (LD) biased
with current denoted by $I_{\mathrm{D}}$. A 10:90 OC, and two isolators
[labeled by arrows in the sketch shown in Fig. \ref{FigSetup}(b)], are
integrated into the fiber loop. The output port of the 10:90 OC is connected
to the POSA setup shown in Fig. \ref{FigSOP}(b).

\begin{figure}[tbp]
\begin{center}
\includegraphics[width=3.2in,keepaspectratio]{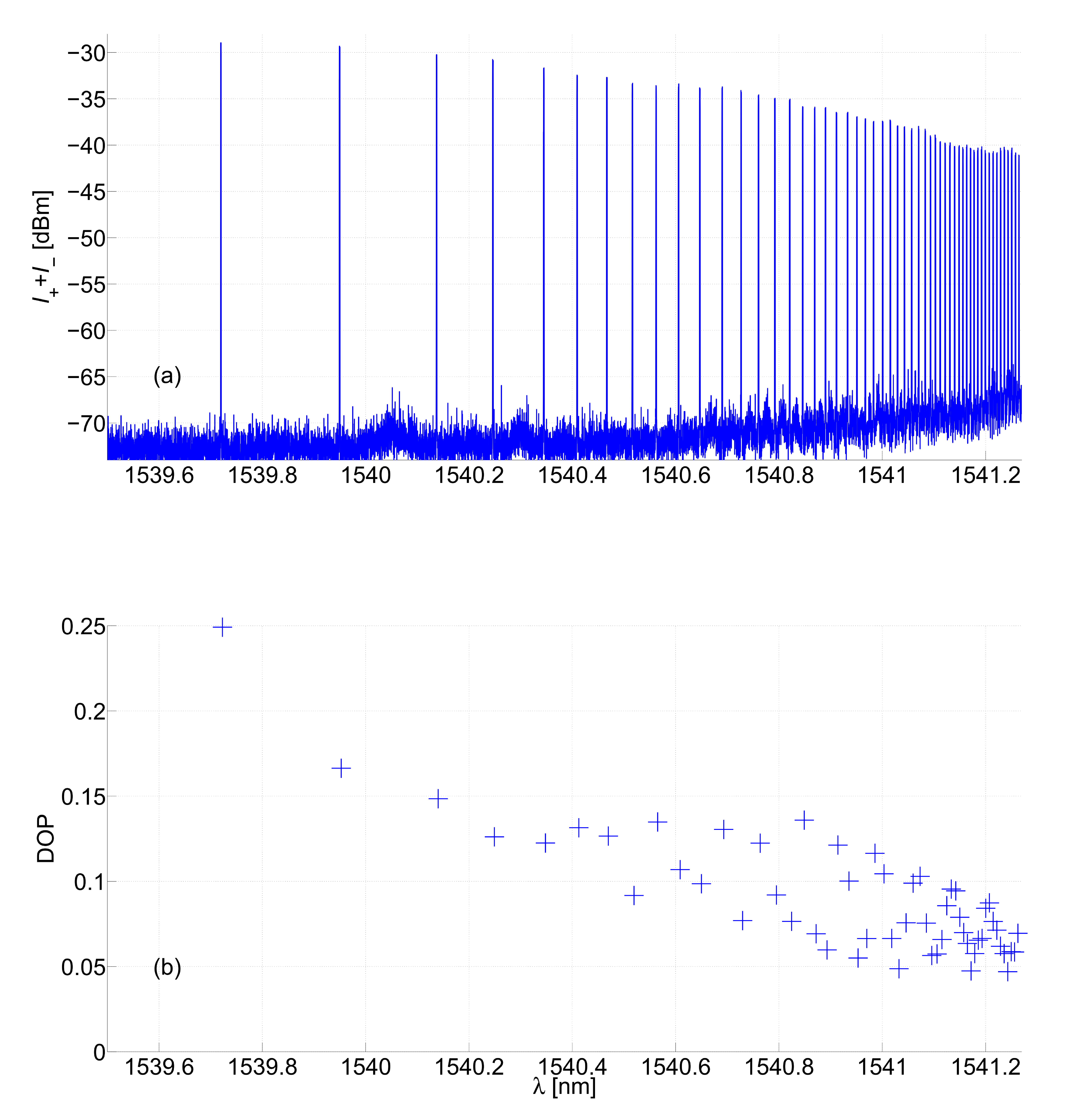}
\end{center}
\caption{{}USOC intensity (a) and DOP (b). Diode current is $0.2\func{A}$,
and EDF temperature is $3.2\func{K}$.}
\label{FigUSOC_DOP}
\end{figure}

The plot in Fig. \ref{FigUSOC_DOP}(a) shows the total POSA signal intensity $%
I_{+}+I_{-}$ as a function of the TL2 wavelength $\lambda$ in the band where
USOC is formed ($1539.7 - 1541.5 \func{nm}$). The underlying mechanism responsible for USOC formation has
remained mainly unknown \cite{Buks_128591,Buks_2951}. POSA is employed
to determine the DOP of each USOC peak. The result, which is shown in Fig. %
\ref{FigUSOC_DOP}(b), reveals that the first (shortest wavelength) USOC peak
has the largest DOP. Note that commonly in fiber lasers DOP is below about $%
0.15$, unless a polarization-maintaining fiber is used \cite{Liem_CMS4}.
Further study is needed to explore how higher USOC DOP, which is needed for
some applications, can be achieved.

\section{Discussion and summary}

A variety of POSA configurations have been developed and optimized for specific applications \cite{trujillo2002astrophysical}. Commonly, wavelength separation in POSA instruments is achieved by integrating either a diffraction grating or a tunable optical filter. For some applications, in fields including astronomy and mineralogy analysis \cite{Belyaev_25980}, these methods provide sufficiently high spectral resolution. However, a much higher spectral resolution can be achieved by implementing the method of coherent heterodyne detection \cite{Yao_17854,Kim_351,Kim_317}.

Our coherent POSA setup operates in the entire C band ($1530-1565 \func{nm}$), and it has dynamic range of about 70 dB. It allows SOP measurement with spectral resolution of $5 \func{MHz}$ (corresponding
wavelength resolution of $0.04 \func{pm}$ in the telecom band). SOP can be
alternatively measured by combining a polarimeter with a tunable optical
filter. The filter-based method is simpler to implement, however, its
spectral resolution is limited by the ratio between the filter's FSR and finesse. Filter-based SOP
measurements of a driven FMSR have been reported in \cite%
{Haigh_143601,Haigh_133602}. In that study, the FMSR resonance frequency $%
\omega _{\mathrm{m}}/\left(2\pi\right)$ was tuned close to $7 \func{GHz}$,
and a scanning Fabry–Pérot etalon having FSR of $10 \func{GHz}
$ (corresponding wavelength spacing of $80 \func{pm}$ in the telecom
band) was employed as a tunable filter \cite{Haigh_143601,Haigh_133602}.
Accurate SOP measurements of a driven FMSR can be obtained provided that the
filter is capable of resolving individual sidebands. This task is
challenging for a driven FMSR, because the power carried by both Stokes and
anti-Stokes optical sidebands is commonly at least 3 orders of magnitude
smaller than the power carried by the pump tone [see Fig. \ref{FigFSR}(2)].
Thus, for SOP measurements of a driven FMSR, the much higher spectral
resolution offered by the coherent POSA significantly improves accuracy. For
the USOC, as can be seen from Fig. \ref{FigUSOC_DOP}(a), the spacing between
peaks is far too small to practically allow any accurate SOP measurements using
the filter-based method.

Future work will be devoted to optimize some applications, that are based on
the optical systems under-study. For the FMSR modulator, ways to further increase the asymmetry parameter $\zeta _{\mathbf{%
a}}$ will be explored to allow more efficient optical communication based on
SSM. Moreover, ways to enhance USOC DOP will be investigated in order to
open the way for some novel applications that require a multi-mode lasing
source having high level of intermode coherency.

\bibliographystyle{ieeepes}
\bibliography{Eyal_Bib}

\begin{thebibliography}{10}

\bibitem{goldstein2017polarized}
Dennis~H Goldstein,
\newblock {\em Polarized light},
\newblock CRC press, 2017.

\bibitem{Yao_17854}
X~Steve Yao, Bo~Zhang, Xiaojun Chen, and Alan~E Willner,
\newblock ``Real-time optical spectrum analysis of a light source using a
  polarimeter'',
\newblock {\em Optics Express}, vol. 16, no. 22, pp. 17854--17863, 2008.

\bibitem{Kim_351}
Eunha Kim, Digant Dave, and Thomas~E Milner,
\newblock ``Fiber-optic spectral polarimeter using a broadband swept laser
  source'',
\newblock {\em Optics communications}, vol. 249, no. 1-3, pp. 351--356, 2005.

\bibitem{Kim_317}
HK~Kim, SK~Kim, HG~Park, and Byoung~Yoon Kim,
\newblock ``Polarimetric fiber laser sensors'',
\newblock {\em Optics letters}, vol. 18, no. 4, pp. 317--319, 1993.

\bibitem{Baney_355}
Douglas~M Baney, Bogdan Szafraniec, and Ali Motamedi,
\newblock ``Coherent optical spectrum analyzer'',
\newblock {\em Ieee Photonics Technology Letters}, vol. 14, no. 3, pp.
  355--357, 2002.

\bibitem{Feng_348}
Kunpeng Feng, Jiwen Cui, Hong Dang, Weidong Wu, Xun Sun, Xuelin Jiang, and
  Jiubin Tan,
\newblock ``An optoelectronic equivalent narrowband filter for high resolution
  optical spectrum analysis'',
\newblock {\em Sensors}, vol. 17, no. 2, pp. 348, 2017.

\bibitem{Dang_4430}
Hong Dang, Huanhuan Liu, Linqi Cheng, Yuqi Tian, Jinna Chen, Kunpeng Feng,
  Jiwen Cui, and Perry~Ping Shum,
\newblock ``Deconvolutional suppression of resolution degradation in coherent
  optical spectrum analyzer'',
\newblock {\em Journal of Lightwave Technology}, vol. 13, pp. 4430, 2023.

\bibitem{Haigh_133602}
JA~Haigh, Andreas Nunnenkamp, AJ~Ramsay, and AJ~Ferguson,
\newblock ``Triple-resonant brillouin light scattering in magneto-optical
  cavities'',
\newblock {\em Physical review letters}, vol. 117, no. 13, pp. 133602, 2016.

\bibitem{Chai_820}
Cheng-Zhe Chai, Zhen Shen, Yan-Lei Zhang, Hao-Qi Zhao, Guang-Can Guo,
  Chang-Ling Zou, and Chun-Hua Dong,
\newblock ``Single-sideband microwave-to-optical conversion in high-q
  ferrimagnetic microspheres'',
\newblock {\em Photonics Research}, vol. 10, no. 3, pp. 820--827, 2022.

\bibitem{Buks_128591}
Eyal Buks,
\newblock ``Low temperature spectrum of a fiber loop laser'',
\newblock {\em Physics Letters A}, vol. 458, pp. 128591, 2023.

\bibitem{Potton_717}
Richard~J Potton,
\newblock ``Reciprocity in optics'',
\newblock {\em Reports on Progress in Physics}, vol. 67, no. 5, pp. 717, 2004.

\bibitem{Rameshti_1}
Babak~Zare Rameshti, Silvia~Viola Kusminskiy, James~A Haigh, Koji Usami, Dany
  Lachance-Quirion, Yasunobu Nakamura, Can-Ming Hu, Hong~X Tang, Gerrit~EW
  Bauer, and Yaroslav~M Blanter,
\newblock ``Cavity magnonics'',
\newblock {\em Physics Reports}, vol. 979, pp. 1--61, 2022.

\bibitem{Kusminskiy_299}
Silvia~Viola Kusminskiy,
\newblock ``Cavity optomagnonics'',
\newblock in {\em Optomagnonic Structures: Novel Architectures for Simultaneous
  Control of Light and Spin Waves}, pp. 299--353. World Scientific, 2021.

\bibitem{Zhu_2012_11119}
Na~Zhu, Xufeng Zhang, Xu~Han, Chang-Ling Zou, and Hong~X Tang,
\newblock ``Inverse faraday effect in an optomagnonic waveguide'',
\newblock {\em arXiv:2012.11119}, 2020.

\bibitem{Juraschek_094407}
Dominik~M Juraschek, Derek~S Wang, and Prineha Narang,
\newblock ``Sum-frequency excitation of coherent magnons'',
\newblock {\em Physical Review B}, vol. 103, no. 9, pp. 094407, 2021.

\bibitem{Bittencourt_014409}
VASV Bittencourt, I~Liberal, and S~Viola Kusminskiy,
\newblock ``Light propagation and magnon-photon coupling in optically
  dispersive magnetic media'',
\newblock {\em Physical Review B}, vol. 105, no. 1, pp. 014409, 2022.

\bibitem{Zhang_123605}
Xufeng Zhang, Na~Zhu, Chang-Ling Zou, and Hong~X Tang,
\newblock ``Optomagnonic whispering gallery microresonators'',
\newblock {\em Physical review letters}, vol. 117, no. 12, pp. 123605, 2016.

\bibitem{VanderZiel_190}
JP~Van~der Ziel, Peter~S Pershan, and LD~Malmstrom,
\newblock ``Optically-induced magnetization resulting from the inverse faraday
  effect'',
\newblock {\em Physical review letters}, vol. 15, no. 5, pp. 190, 1965.

\bibitem{Zheng_2303_16313}
Shasha Zheng, Zhenyu Wang, Yipu Wang, Fengxiao Sun, Qiongyi He, Peng Yan, and
  HY~Yuan,
\newblock ``Tutorial: Nonlinear magnonics'',
\newblock {\em arXiv:2303.16313}, 2023.

\bibitem{Stancil_Spin}
Daniel~D Stancil and Anil Prabhakar,
\newblock {\em Spin waves},
\newblock Springer, 2009.

\bibitem{Oraevsky_377}
Anatolii~N Oraevsky,
\newblock ``Whispering-gallery waves'',
\newblock {\em Quantum electronics}, vol. 32, no. 5, pp. 377, 2002.

\bibitem{Schunk_30795}
Gerhard Schunk, Josef~U F{\"u}rst, Michael F{\"o}rtsch, Dmitry~V Strekalov,
  Ulrich Vogl, Florian Sedlmeir, Harald~GL Schwefel, Gerd Leuchs, and Christoph
  Marquardt,
\newblock ``Identifying modes of large whispering-gallery mode resonators from
  the spectrum and emission pattern'',
\newblock {\em Optics express}, vol. 22, no. 25, pp. 30795--30806, 2014.

\bibitem{Gorodetsky_33}
Michael~L Gorodetsky and Aleksey~E Fomin,
\newblock ``Geometrical theory of whispering-gallery modes'',
\newblock {\em IEEE Journal of Selected Topics in Quantum Electronics}, vol.
  12, no. 1, pp. 33--39, 2006.

\bibitem{Walker_390}
Laurence~R Walker,
\newblock ``Magnetostatic modes in ferromagnetic resonance'',
\newblock {\em Physical Review}, vol. 105, no. 2, pp. 390, 1957.

\bibitem{Osada_103018}
A~Osada, A~Gloppe, Y~Nakamura, and K~Usami,
\newblock ``Orbital angular momentum conservation in brillouin light scattering
  within a ferromagnetic sphere'',
\newblock {\em New Journal of Physics}, vol. 20, no. 10, pp. 103018, 2018.

\bibitem{Osada_223601}
A~Osada, R~Hisatomi, A~Noguchi, Y~Tabuchi, R~Yamazaki, K~Usami, M~Sadgrove,
  R~Yalla, M~Nomura, and Y~Nakamura,
\newblock ``Cavity optomagnonics with spin-orbit coupled photons'',
\newblock {\em Physical review letters}, vol. 116, no. 22, pp. 223601, 2016.

\bibitem{Sharma_094412}
Sanchar Sharma, Yaroslav~M Blanter, and Gerrit~EW Bauer,
\newblock ``Light scattering by magnons in whispering gallery mode cavities'',
\newblock {\em Physical Review B}, vol. 96, no. 9, pp. 094412, 2017.

\bibitem{Sharma_087205}
Sanchar Sharma, Yaroslav~M Blanter, and Gerrit~EW Bauer,
\newblock ``Optical cooling of magnons'',
\newblock {\em Physical review letters}, vol. 121, no. 8, pp. 087205, 2018.

\bibitem{Almpanis_184406}
Evangelos Almpanis,
\newblock ``Dielectric magnetic microparticles as photomagnonic cavities:
  Enhancing the modulation of near-infrared light by spin waves'',
\newblock {\em Physical Review B}, vol. 97, no. 18, pp. 184406, 2018.

\bibitem{Zivieri_165406}
R~Zivieri, P~Vavassori, L~Giovannini, F~Nizzoli, Eric~E Fullerton,
  M~Grimsditch, and V~Metlushko,
\newblock ``Stokes--anti-stokes brillouin intensity asymmetry of spin-wave
  modes in ferromagnetic films and multilayers'',
\newblock {\em Physical Review B}, vol. 65, no. 16, pp. 165406, 2002.

\bibitem{Desormiere_379}
BERNARD Desormiere and HENRI Le~Gall,
\newblock ``Interaction studies of a laser light with spin waves and
  magnetoelastic waves propagating in a yig bar'',
\newblock {\em IEEE Transactions on Magnetics}, vol. 8, no. 3, pp. 379--381,
  1972.

\bibitem{Liu_3698}
Zeng-Xing Liu, Bao Wang, Hao Xiong, and Ying Wu,
\newblock ``Magnon-induced high-order sideband generation'',
\newblock {\em Optics Letters}, vol. 43, no. 15, pp. 3698--3701, 2018.

\bibitem{Zhu_1291}
Na~Zhu, Xufeng Zhang, Xu~Han, Chang-Ling Zou, Changchun Zhong, Chiao-Hsuan
  Wang, Liang Jiang, and Hong~X Tang,
\newblock ``Waveguide cavity optomagnonics for microwave-to-optics
  conversion'',
\newblock {\em Optica}, vol. 7, no. 10, pp. 1291--1297, 2020.

\bibitem{Li_040344}
Jie Li, Yi-Pu Wang, Wei-Jiang Wu, Shi-Yao Zhu, and JQ~You,
\newblock ``Quantum network with magnonic and mechanical nodes'',
\newblock {\em PRX Quantum}, vol. 2, no. 4, pp. 040344, 2021.

\bibitem{Nayak_193905}
Banoj~Kumar Nayak and Eyal Buks,
\newblock ``Polarization-selective magneto-optical modulation'',
\newblock {\em Journal of Applied Physics}, vol. 132, no. 19, pp. 193905, 2022.

\bibitem{Liu_5452}
Zeng-Xing Liu and Hao Xiong,
\newblock ``Magnon laser based on brillouin light scattering'',
\newblock {\em Optics Letters}, vol. 45, no. 19, pp. 5452--5455, 2020.

\bibitem{Sandercock_1729}
JR~Sandercock and W~Wettling,
\newblock ``Light scattering from thermal acoustic magnons in yttrium iron
  garnet'',
\newblock {\em Solid State Communications}, vol. 13, no. 10, pp. 1729--1732,
  1973.

\bibitem{Hu_307}
HL~Hu and FR~Morgenthaler,
\newblock ``Strong infrared-light scattering from coherent spin waves in
  yttrium iron garnet'',
\newblock {\em Applied Physics Letters}, vol. 18, no. 7, pp. 307--310, 1971.

\bibitem{Ghasemian_12757}
E~Ghasemian,
\newblock ``Dissipative dynamics of optomagnonic nonclassical features via
  anti-stokes optical pulses: squeezing, blockade, anti-correlation, and
  entanglement'',
\newblock {\em Scientific Reports}, vol. 13, no. 1, pp. 12757, 2023.

\bibitem{Kusminskiy_033821}
Silvia~Viola Kusminskiy, Hong~X Tang, and Florian Marquardt,
\newblock ``Coupled spin-light dynamics in cavity optomagnonics'',
\newblock {\em Physical Review A}, vol. 94, no. 3, pp. 033821, 2016.

\bibitem{Wettling_211}
W~Wettling, MG~Cottam, and JR~Sandercock,
\newblock ``The relation between one-magnon light scattering and the complex
  magneto-optic effects in yig'',
\newblock {\em Journal of Physics C: Solid State Physics}, vol. 8, no. 2, pp.
  211, 1975.

\bibitem{cottam1986light}
Michael~G Cottam and David~J Lockwood,
\newblock {\em Light scattering in magnetic solids},
\newblock Wiley New York, 1986.

\bibitem{Liu_060405}
Tianyu Liu, Xufeng Zhang, Hong~X Tang, and Michael~E Flatt{\'e},
\newblock ``Optomagnonics in magnetic solids'',
\newblock {\em Physical Review B}, vol. 94, no. 6, pp. 060405, 2016.

\bibitem{Haigh_143601}
JA~Haigh, A~Nunnenkamp, and AJ~Ramsay,
\newblock ``Polarization dependent scattering in cavity optomagnonics'',
\newblock {\em Physical Review Letters}, vol. 127, no. 14, pp. 143601, 2021.

\bibitem{Hisatomi_207401}
R~Hisatomi, A~Noguchi, R~Yamazaki, Y~Nakata, A~Gloppe, Y~Nakamura, and K~Usami,
\newblock ``Helicity-changing brillouin light scattering by magnons in a
  ferromagnetic crystal'',
\newblock {\em Physical Review Letters}, vol. 123, no. 20, pp. 207401, 2019.

\bibitem{Hisatomi_174427}
Ryusuke Hisatomi, Alto Osada, Yutaka Tabuchi, Toyofumi Ishikawa, Atsushi
  Noguchi, Rekishu Yamazaki, Koji Usami, and Yasunobu Nakamura,
\newblock ``Bidirectional conversion between microwave and light via
  ferromagnetic magnons'',
\newblock {\em Physical Review B}, vol. 93, no. 17, pp. 174427, 2016.

\bibitem{Wu_023711}
Wei-Jiang Wu, Yi-Pu Wang, Jin-Ze Wu, Jie Li, and JQ~You,
\newblock ``Remote magnon entanglement between two massive ferrimagnetic
  spheres via cavity optomagnonics'',
\newblock {\em Physical Review A}, vol. 104, no. 2, pp. 023711, 2021.

\bibitem{lazzarini2008asymmetric}
Victor Lazzarini, Joseph Timoney, and Thomas Lysaght,
\newblock ``Asymmetric-spectra methods for adaptive fm synthesis'',
\newblock {\em MURAL - Maynooth University Research Archive Library}, 2008.

\bibitem{Li_1}
Wei Li, Wen~Ting Wang, Li~Xian Wang, and Ning~Hua Zhu,
\newblock ``Optical vector network analyzer based on single-sideband modulation
  and segmental measurement'',
\newblock {\em IEEE Photonics Journal}, vol. 6, no. 2, pp. 1--8, 2014.

\bibitem{Shimotsu_364}
S~Shimotsu, S~Oikawa, T~Saitou, N~Mitsugi, K~Kubodera, T~Kawanishi, and
  M~Izutsu,
\newblock ``Single side-band modulation performance of a linbo 3 integrated
  modulator consisting of four-phase modulator waveguides'',
\newblock {\em IEEE Photonics Technology Letters}, vol. 13, no. 4, pp.
  364--366, 2001.

\bibitem{Buks_2951}
Eyal Buks,
\newblock ``Intermode coupling in a fiber loop laser at low temperatures'',
\newblock {\em Journal of Lightwave Technology}, vol. 42, pp. 2951, 2024.

\bibitem{Onbasli_1}
Mehmet~C Onbasli, Luk{\'a}{\v{s}} Beran, Martin Zahradn{\'\i}k, Miroslav
  Ku{\v{c}}era, Roman Anto{\v{s}}, Jan Mistr{\'\i}k, Gerald~F Dionne, Martin
  Veis, and Caroline~A Ross,
\newblock ``Optical and magneto-optical behavior of cerium yttrium iron garnet
  thin films at wavelengths of 200--1770 nm'',
\newblock {\em Scientific reports}, vol. 6, 2016.

\bibitem{Fletcher_687}
PC~Fletcher and RO~Bell,
\newblock ``Ferrimagnetic resonance modes in spheres'',
\newblock {\em Journal of Applied Physics}, vol. 30, no. 5, pp. 687--698, 1959.

\bibitem{sharma2019cavity}
S~Sharma,
\newblock {\em Cavity optomagnonics: Manipulating magnetism by light},
\newblock PhD thesis, Delft University of Technology, 2019.

\bibitem{Jin_thesis}
Tien~Liang Jin,
\newblock ``Design of a yig-tuned oscillator'',
\newblock {\em New Jersey Institute of Technology}, 1974.

\bibitem{Freiser_152}
Ml~Freiser,
\newblock ``A survey of magnetooptic effects'',
\newblock {\em IEEE Transactions on magnetics}, vol. 4, no. 2, pp. 152--161,
  1968.

\bibitem{Boardman_197}
Allan~D Boardman and Ming Xie,
\newblock ``Magneto-optics: a critical review'',
\newblock {\em Introduction to Complex Mediums for Optics and
  Electromagnetics}, vol. 123, pp. 197, 2003.

\bibitem{Boardman_388}
Allan~D Boardman and Larry Velasco,
\newblock ``Gyroelectric cubic-quintic dissipative solitons'',
\newblock {\em IEEE Journal of selected topics in quantum electronics}, vol.
  12, no. 3, pp. 388--397, 2006.

\bibitem{Wood_1038}
DL~Wood and JP~Remeika,
\newblock ``Effect of impurities on the optical properties of yttrium iron
  garnet'',
\newblock {\em Journal of Applied Physics}, vol. 38, no. 3, pp. 1038--1045,
  1967.

\bibitem{Prokhorov_339}
Aleksandr~Mikhailovich Prokhorov and Smolenskii,
\newblock ``Optical phenomena in thin-film magnetic waveguides and their
  technical application'',
\newblock {\em Soviet Physics Uspekhi}, vol. 27, no. 5, pp. 339, 1984.

\bibitem{prabhakar2009spin}
Anil Prabhakar and Daniel~D Stancil,
\newblock {\em Spin waves: Theory and applications}, vol.~5,
\newblock Springer, 2009.

\bibitem{Buks_QMLN}
Eyal Buks,
\newblock {\em Quantum mechanics - Lecture Notes},
\newblock http://buks.net.technion.ac.il/teaching/, 2024.

\bibitem{Buks_L051001}
Eyal Buks,
\newblock ``Tunable multimode lasing in a fiber ring'',
\newblock {\em Physical Review Applied}, vol. 19, no. 5, pp. L051001, 2023.

\bibitem{Yamashita_1298}
S~Yamashita and K~Hotate,
\newblock ``Multiwavelength erbium-doped fibre laser using intracavity etalon
  and cooled by liquid nitrogen'',
\newblock {\em Electronics Letters}, vol. 32, no. 14, pp. 1298--1299, 1996.

\bibitem{Liu_102988}
Haochong Liu, Wei He, Yantao Liu, Yunhui Dong, and Lianqing Zhu,
\newblock ``Erbium-doped fiber laser based on femtosecond laser inscribed fbg
  through fiber coating for strain sensing in liquid nitrogen environment'',
\newblock {\em Optical Fiber Technology}, vol. 72, pp. 102988, 2022.

\bibitem{Lopez_085401}
J~Lopez, H~Kerbertt, M~Plata, E~Hernandez, and S~Stepanov,
\newblock ``Two-wave mixing in erbium-doped-fibers with spectral-hole burning
  at 77k'',
\newblock {\em Journal of Optics}, vol. 22, no. 8, pp. 085401, 2020.

\bibitem{Le_3611}
Julien Le~Gou{\"e}t, J{\'e}r{\'e}my Oudin, Philippe Perrault, Alaeddine Abbes,
  Alice Odier, and Aliz{\'e}e Dubois,
\newblock ``On the effect of low temperatures on the maximum output power of a
  coherent erbium-doped fiber amplifier'',
\newblock {\em Journal of Lightwave Technology}, vol. 37, no. 14, pp.
  3611--3619, 2019.

\bibitem{Wei_2209_00802}
Shi-Hai Wei, Bo~Jing, Xue-Ying Zhang, Jin-Yu Liao, Hao Li, Li-Xing You, Zhen
  Wang, You Wang, Guang-Wei Deng, Hai-Zhi Song, et~al.,
\newblock ``Storage of 1650 modes of single photons at telecom wavelength'',
\newblock {\em arXiv:2209.00802}, 2022.

\bibitem{Ortu_035024}
Antonio Ortu, Jelena~V Rakonjac, Adrian Holz{\"a}pfel, Alessandro Seri, Samuele
  Grandi, Margherita Mazzera, Hugues de~Riedmatten, and Mikael Afzelius,
\newblock ``Multimode capacity of atomic-frequency comb quantum memories'',
\newblock {\em Quantum Science and Technology}, vol. 7, no. 3, pp. 035024,
  2022.

\bibitem{Liu_2201_03692}
Duan-Cheng Liu, Pei-Yun Li, Tian-Xiang Zhu, Liang Zheng, Jian-Yin Huang,
  Zong-Quan Zhou, Chuan-Feng Li, and Guang-Can Guo,
\newblock ``On-demand storage of photonic qubits at telecom wavelengths'',
\newblock {\em arXiv:2201.03692}, 2022.

\bibitem{Veissier_195138}
Lucile Veissier, Mohsen Falamarzi, Thomas Lutz, Erhan Saglamyurek, Charles~W
  Thiel, Rufus~L Cone, and Wolfgang Tittel,
\newblock ``Optical decoherence and spectral diffusion in an erbium-doped
  silica glass fiber featuring long-lived spin sublevels'',
\newblock {\em Physical Review B}, vol. 94, no. 19, pp. 195138, 2016.

\bibitem{Shafiei_F2A}
Sara Shafiei, Erhan Saglamyurek, and Daniel Oblak,
\newblock ``Hour-long decay-time of erbium spins in an optical fiber at
  milli-kelvin temperatures'',
\newblock in {\em Quantum Information and Measurement}. Optical Society of
  America, 2021, pp. F2A--4.

\bibitem{Liem_CMS4}
A~Liem, J~Limpert, T~Schreiber, M~Reich, H~Zellmer, A~Tunnermann, A~Carter, and
  K~Tankala,
\newblock ``High power linearly polarized fiber laser'',
\newblock in {\em Conference on Lasers and Electro-Optics}. Optica Publishing
  Group, 2004, p. CMS4.

\bibitem{trujillo2002astrophysical}
Javier Trujillo-Bueno, Fernando Moreno-Insertis, and Francisco S{\'a}nchez,
\newblock {\em Astrophysical spectropolarimetry},
\newblock Cambridge University Press, 2002.

\bibitem{Belyaev_25980}
Denis~A Belyaev, Konstantin~B Yushkov, Sergey~P Anikin, Yuri~S Dobrolenskiy,
  Aleksander Laskin, Sergey~N Mantsevich, Vladimir~Ya Molchanov, Sergey~A
  Potanin, and Oleg~I Korablev,
\newblock ``Compact acousto-optic imaging spectro-polarimeter for mineralogical
  investigations in the near infrared'',
\newblock {\em Optics Express}, vol. 25, no. 21, pp. 25980--25991, 2017.

\end{thebibliography}

\end{document}